\documentclass[prl,amsmath,showpacs,twocolumn]{revtex4}
\usepackage{graphicx}
\usepackage{multirow}
\usepackage{amssymb}

\begin{document}

\title{Magnetoelectric Response of Multiferroic BiFeO$_3$ and Related
  Materials}

\author{Jacek C. Wojde\l\ and Jorge \'I\~niguez}

\affiliation{Institut de Ci\`encia de Materials de Barcelona
(ICMAB-CSIC), Campus UAB, 08193 Bellaterra, Spain}

\begin{abstract}
  We present a first-principles scheme for computing the magnetoelectric
  response of multiferroics. We apply our method to BiFeO$_3$ (BFO) and
  related compounds in which Fe is substituted by other magnetic species. We
  show that under certain relevant conditions -- i.e., in absence of
  incommensurate spin modulation, as in BFO thin films and some BFO-based
  solid solutions -- these materials display a large linear magnetoelectric
  response. Our calculations reveal the atomistic origin of the coupling and
  allow us to identify the most promising strategies to enhance it.
\end{abstract}





\maketitle

Magnetoelectric (ME) multiferroics present coupled electric and magnetic
orders~\cite{fiebig05}, which may allow the development of a variety of
magnetic devices, from memories to spin filters, whose behavior would be {\sl
  switchable} by application of a voltage. The experimental search for robust
room-temperature ($T_{\rm room}$) multiferroics is proving a major
challenge. The first-principles contribution to this effort is quickly
increasing, as shown by recent predictions of new
materials~\cite{baettig05,picozzi07,bhattacharjee09} and novel ME coupling
mechanisms~\cite{rondinelli08,fennie08,delaney09}. Yet, many key issues remain
to be addressed theoretically, as even the ME response of the most promising
systems (e.g., thin films of BiFeO$_3$~\cite{catalan09}) still needs to be
characterized and understood in detail. Here we describe a method for {\sl ab
  initio} computations of the ME response of multiferroics, tackling the
strain-mediated contributions that typically exist in these materials.  We
apply our method to BiFeO$_3$ and related compounds.

{\sl Methodology}.-- The response of a linear magnetoelectric that is not
multiferroic -- and is thus paraelectric -- can be split in two
contributions~\cite{iniguez08}: a purely electronic one and an ionic one. The
ionic part accounts for the structural response to an electric (magnetic)
field and the resulting change of magnetization (polarization). When the
linear magnetoelectric is multiferroic -- and thus ferroelectric -- there is a
third contribution associated to the piezoelectric response to the electric
field. Indeed, it can be shown that all {\sl commensurate} \cite{fn:sym}
multiferroics must be piezomagnetic, as the symmetries that preclude
piezomagnetism are broken in such systems~\cite{fn:piezo}. Hence, the
combination of piezoelectricity and piezomagnetism in multiferroics results in
a strain-mediated contribution to the linear ME response.

In order to model such effects, we have generalized to the
magnetoelectric case the formalism that Wu, Vanderbilt, and
Hamann~\cite{wu05} (WVH in the following) introduced for a systematic
{\sl ab initio} treatment of dielectric and piezoelectric responses.
Let us consider how the energy of a multiferroic crystal varies as a
function of the following perturbations: displacements $u_m$ of atoms
away from their equilibrium positions, homogeneous strains $\eta_j$,
and applied electric ($\cal{E}_{\alpha}$) and magnetic
($\cal{H}_{\mu}$) fields. Here, $m$ is a composite label that runs
over the atoms in the unit cell and displacement directions, $j$ runs
from 1 to 6 in Voigt notation, and $\alpha$ and $\mu$ are Cartesian
directions. We write the energy per undeformed unit cell volume of the
multiferroic crystal, expanded to second order around its zero-field
equilibrium structure, as
\begin{equation}
\label{eq:e}
{\small
\begin{split}
E& = E_0 + A_{\alpha} {\cal E}_{\alpha} 
        + A_{\mu} {\cal H}_{\mu} + \frac{1}{2} B_{mn} u_m u_n
    + \frac{1}{2} B_{\alpha\beta} {\cal E}_{\alpha} {\cal E}_{\beta} \\
&  + \frac{1}{2} B_{\mu\nu} {\cal H}_{\mu} {\cal H}_{\nu}  +
    \frac{1}{2} B_{jk} \eta_j \eta_k  + B_{m\alpha} u_m {\cal E}_{\alpha} +
    B_{m\mu} u_m {\cal H}_{\mu}  \\
& + B_{mj} u_m \eta_j +  B_{\alpha\mu} {\cal E}_{\alpha}  {\cal
      H}_{\mu} + B_{\alpha j}
    {\cal E}_{\alpha} \eta_j + B_{\mu j} {\cal H}_{\mu} \eta_j \, ,
\end{split}
}
\end{equation}
assuming summation over repeated indexes. The equilibrium condition
implies the terms linear in $u_m$ and $\eta_j$ are zero; the term
$A_{\alpha}$=$-\cal{P}^{\rm S}_{\alpha}$ ($A_{\mu}$=$-\cal{M}^{\rm
  S}_{\mu}$) is the spontaneous polarization (magnetization). The
second derivatives of Eq.~(\ref{eq:e}) correspond to well-known
physical quantities, such as the force constant matrix
$K_{mn}=\Omega_0 B_{mn}$, the frozen-ion elastic tensor
$\bar{C}_{jk}=B_{jk}$, the Born dynamical effective charges
$Z_{m\alpha}=-\Omega_0 B_{m\alpha}$, the force-response internal
strain tensor $\Lambda_{mj}=-\Omega_0 B_{mj}$, or the frozen-ion
piezoelectric stress tensor $\bar{e}_{\alpha j}=-B_{\alpha j}$, where
$\Omega_0$ is the unit cell volume and the bar stands for {\sl
  frozen-ion}. Additionally, the purely electronic ME response is
\begin{equation}
{\small
\hat{\bar{\alpha}}_{\alpha \mu} = - \frac{\partial^2 E}{\partial
  {\cal E}_{\alpha} \partial {\cal H}_{\mu} } \Bigr\rvert_{u,\eta} = 
\frac{\partial\cal{P}_{\alpha}}{\partial\cal{H}_{\mu}}\Bigr\rvert_{u,\eta} =
- B_{\alpha\mu} \, ,
}
\end{equation}
where the hat stands for {\sl frozen-cell}. The frozen-ion piezomagnetic
stress tensor is
\begin{equation}
{\small
\bar{h}_{\mu j} = - \frac{\partial^2 E}{\partial {\cal H}_{\mu} \partial
  \eta_j } \Bigr\rvert_{u,{\cal E}} = 
\frac{\partial\cal{M}_{\mu}}{\partial\eta_j}\Bigr\rvert_{u,{\cal E}} = 
- B_{\mu j} \, .
}
\end{equation}
Lastly, the magnetization change driven by an atomic displacement is
\begin{equation}
{\small
\zeta_{m\mu} = - \Omega_0 \frac{\partial^2 E}{\partial
  u_m \partial {\cal H}_{\mu} } \Bigr\rvert_{\eta,{\cal E}} = 
- \Omega_0 B_{m\mu} \, .
}
\end{equation}
The tensors describing the response to static fields should account
for the field-induced ionic relaxation. To compute such {\sl
  relaxed-ion} quantities, we introduce the energy $\tilde{E}({\cal
  E},{\cal
  H},\eta)=\min_u E(u,{\cal E},{\cal H},\eta)$, which is obtained from
Eq.~(\ref{eq:e}) by replacing
\begin{equation}
{\small
u_m = - (B)^{-1}_{mn} (B_{n\alpha}{\cal E}_{\alpha} + B_{n\mu} H_{\mu} +
B_{nj} \eta_j) \, .
}
\end{equation}
$\tilde{E}$ can be written in a form analogous to Eq.~(\ref{eq:e}),
with $\tilde{B}$ coefficients that are combinations of the original
$B$'s. By computing the second derivatives of $\tilde{E}$ we can
obtain, for example, the piezomagnetic stress tensor
\begin{equation}
{\small
\begin{split}
h_{\mu j} & = - \frac{\partial^2 \tilde{E}}{\partial {\cal H}_{\mu} \partial
  \eta_j } \Bigr\rvert_{{\cal E}} = 
- \tilde{B}_{\mu j} = - B_{\mu j} + B_{m\mu} (B)^{-1}_{mn} B_{nj} \\
& = \bar{h}_{\mu j} + \Omega_{0}^{-1} \zeta_{m\mu} (K^{-1})_{mn} \Lambda_{nj}
\end{split}
}
\end{equation}
or the frozen-cell ME tensor
\begin{equation}
\label{eq:hatalpha}
{\small
  \hat{\alpha}_{\alpha\mu} = - \frac{\partial^2 \tilde{E}}{\partial
    {\cal E}_{\alpha} \partial {\cal H}_{\mu} }\Bigr\rvert_{\eta} =
 \hat{\bar{\alpha}}_{\alpha\mu} + \Omega_0^{-1} Z_{m\alpha}
  (K^{-1})_{mn} \zeta_{n\mu} \, .
}
\end{equation}
Finally, in order to compute the full ME response we go one step
beyond WVH and introduce the energy ${\mathfrak
  E}({\cal E},{\cal H})=\min_{\eta} \tilde{E}({\cal E},{\cal H},\eta)$, from
which we derive
\begin{equation}
\label{eq:totalalpha}
{\small
\begin{split}
\alpha_{\alpha\mu} & = - \frac{\partial^2 {\mathfrak E}}{\partial
    {\cal E}_{\alpha} \partial {\cal H}_{\mu} } = \\
& \hat{\bar{\alpha}}_{\alpha\mu} + \Omega_0^{-1} Z_{m\alpha}
  (K^{-1})_{mn} \zeta_{n\mu} + e_{\alpha j} (C^{-1})_{jk} h_{\mu k} \, .
\end{split}
}
\end{equation}
Note that, except for the purely electronic $\hat{\bar{\alpha}}_{\alpha\mu}$,
all the terms in Eq.~(\ref{eq:totalalpha}) can be obtained from finite
difference calculations that do not require the simulation of applied
fields. Thus, the present formalism brings the study of ME effects in
multiferroics within the scope of the most widely used Density Functional
Theory (DFT) codes.

Finally, let us remark we would like our theory to describe the
so-called {\sl proper} piezoelectric effects. WVH dealt with this
issue by introducing suitably rescaled electric fields and
polarizations~\cite{wu05}. Equivalently, the desired result is
obtained if the $B_{\alpha j}$ coefficients of Eq.~(\ref{eq:e}) are
computed by finite differences as described in
Ref.~\onlinecite{vanderbilt00}. Interestingly, the proper {\sl vs}
improper distinction applies to the piezomagnetic case too, a fact
that has seemingly remained unnoticed. Indeed, we want our {\sl
  proper} $B_{\mu j}$ coefficients to be such that (i) ${\cal
  H}$-induced rotations of the sample do not contribute to the
response and (ii) a mere change in the unit cell volume has no
magnetic effect. The former requirement is easily accomplished by
working only with the symmetric part of the strain tensor, and we
comply with the latter by computing the $B_{\mu j}$ coefficients from
strain-induced changes in the magnetic moment {\sl
  per} cell, as opposed to changes in the
magnetization~\cite{fn:addcomments}.

\begin{figure}[t!]
\raisebox{6cm}{(a)}\
\includegraphics*[width=0.4\columnwidth,viewport=0 -40 390 665]{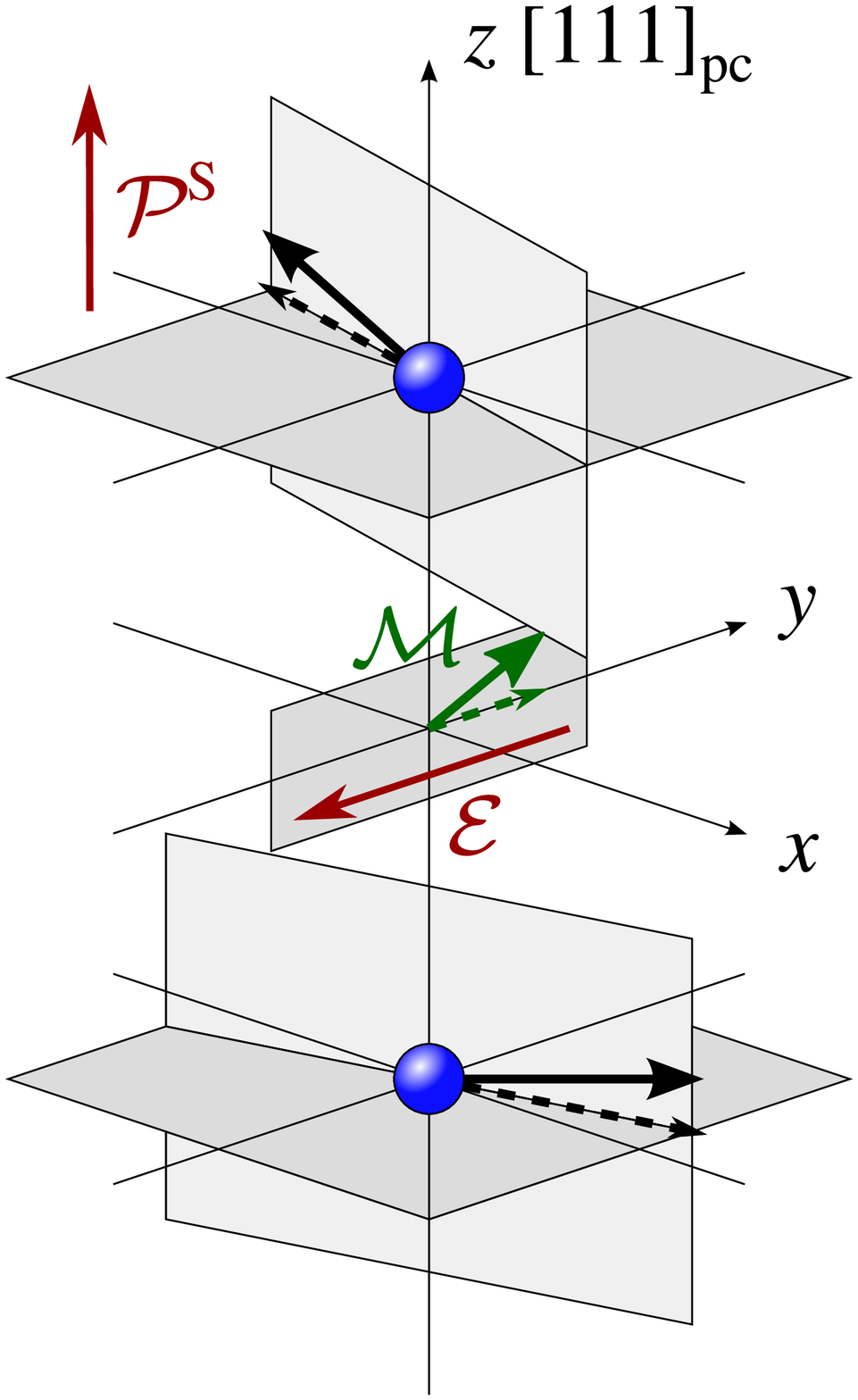}\
\raisebox{6cm}{(b)}\
\includegraphics*[width=0.45\columnwidth,viewport=50 5 420 600]{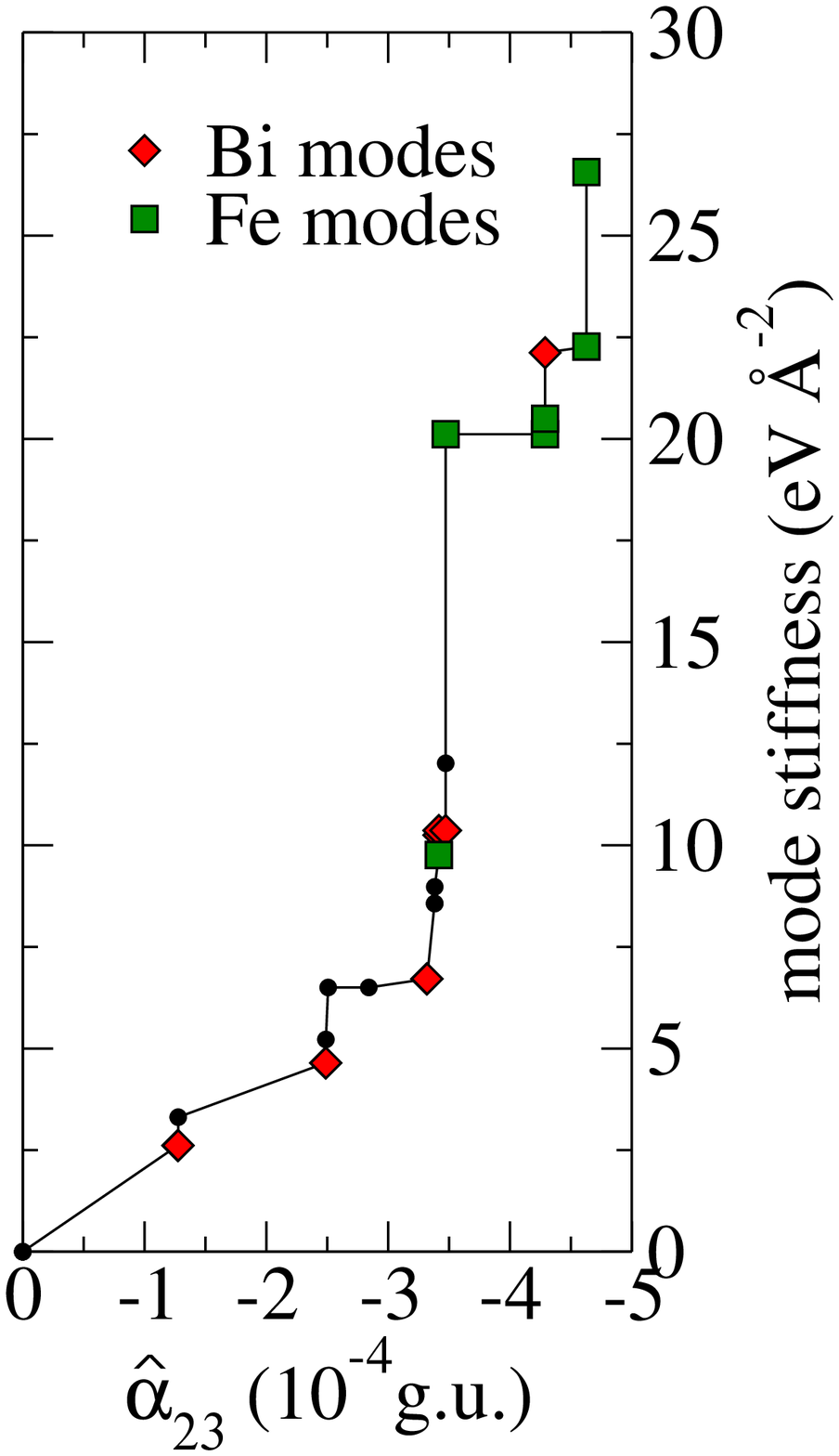}\
\vskip -2mm
\caption{Panel~(a): Sketch of the response associated to $\alpha_{23}$ of
  BFO. Only the two Fe atoms in the usual (rhombohedral) cell of BFO are
  shown. In equilibrium (dashed arrows) the spin of the Fe ions lies on the
  $xy$ plane; the easy axis is $x$ and a small canting results in a net ${\cal
    M}_y^{\rm S}$. When ${\cal E}_y$ is applied we obtain an additional
  canting and a non-zero ${\cal M}_z$ (solid arrows).  Panel~(b): Cumulative
  plot for the $\hat{\alpha}_{23}$ coefficient of BFO (see text). Modes
  dominated by Bi (green diamonds) and Fe (red squares) displacements are
  highlighted.\label{fig1}}
\end{figure}

{\sl Application to BiFeO$_3$ and related materials}.-- We applied our method
to BiFeO$_3$ (BFO), arguably the most promising multiferroic.  We simulated
BFO with the magnetic structure of the systems that are most relevant for
applications, i.e., BFO thin films and solid solutions in which Bi is partly
substituted by a lanthanide to improve the dielectric properties. In such
cases BFO looses the spin cycloid that occurs in bulk
samples~\cite{bai05,bea07,zalesskii03} and presents a canted G-type
antiferromagnetic (AFM) spin arrangement. The magnetic easy axis lies within
the plane perpendicular to the polar direction ($xy$ in Fig.~\ref{fig1}(a),
$z$ being parallel to ${\cal P}^{\rm S}$ and corresponding to the [111]$_{\rm
  pc}$ pseudocubic direction)~\cite{bea07,ederer05}, and the corresponding
magnetic space group allows for a linear ME
response~\cite{fn:spacegroup}. Indeed, there is experimental evidence of such
a linear ME effect~\cite{wang03,jiang07}, but we still lack a detailed and
well-established characterization.

We also studied situations in which, maintaining the 10-atom cell and basic
structure of BFO, Fe is substituted by other magnetic species.  In particular,
we have considered Fe$\rightarrow${\sl Cr}, Fe$\rightarrow${\sl Mn}, and
Fe$\rightarrow${\sl Co} substitutions, as well as Fe$\rightarrow$(Fe,Cr) and
Fe$\rightarrow${\sl (Mn,Ni)} double perovskites~\cite{fn:highspin}. Note that,
except for Bi$_2$FeCrO$_6$, none of the studied substitutions corresponds to
thermodymically stable phases. The purpose of considering such ficticious
compounds (denoted with italics) was to identify chemical trends in the
magnitude of the ME effects.

For the calculations we used the Projector Augmented Wave approach to DFT,
within the so-called ``LDA+$U$'' approximation, as implemented in the {\sl
  Vienna Ab initio Simulation Package}~\cite{fn:calcs}. For all the
compositions studied, we proceeded as follows: Taking the usual atomic and
magnetic structure of rhombohedral BFO as starting point, we relaxed the
system to find a well-defined energy minimum. This energy minimization
included a careful search for the magnetic easy axis. Having identified the
equilibrium state, we computed the coefficients in Eq.~(\ref{eq:e}) by finite
differences. All the calculations were fully self-consistent, allowed for
non-collinear magnetism, and included spin-orbit couplings. We neglected
orbital magnetization as well as the purely electronic ME response
$\mathbf{\hat{\bar{\alpha}}}$, which is likely to be relatively
small~\cite{fn:latticevselectronic}, specially for materials with a
significant structural response to electric fields.

\begin{table}[t!]
\caption{\label{tab1} Top: Frozen-cell and full ME tensors for three
  representative cases. Results given in 10$^{-4}$ Gaussian units
  (g.u.). Magnetic space groups and easy axes are indicated. Bi$_2$FeCrO$_6$
  has nearly $R3$ symmetry, except for a tiny spin canting. Dots indicate
  coefficients that are zero by symmetry. Bottom: Maximum ME response, as
  quantified by square root of largest eigenvalue of
  $\mathbf{\alpha}^{t}\mathbf{\alpha}$, for all considered substitutions.}
\vskip 1mm 
\begin{tabular*}{0.97\columnwidth}{@{\extracolsep{\fill}}cccc}
\hline\hline
& BiFeO$_3$ & Bi$_2$FeCrO$_6$ & {\sl BiCrO$_3$} \\
space group & $Bb'$  & $\sim R3$  & $R3c$ \\
easy axis   & $x$ & $z$ & $z$ \\ 
$\mathbf{\cal M}$($\mu_{B}$/cell) & 0.036${\parallel}y$ & 
0.002${\parallel}y$, -2${\parallel}z$ & 
null \\[1mm]
$\mathbf{\hat{\alpha}}$ &
$\begin{bmatrix}0 & . & . \\ . & 0 & -5 \\ . & 0 & 0 \end{bmatrix}$& 
$\begin{bmatrix}-4 & -12 & . \\ 9 &  -4 & . \\ . &   . & 0 \end{bmatrix}$&
$\begin{bmatrix} . & -4 & . \\  4 & . & . \\ . &   . & .\end{bmatrix}$
\\ [6mm]
$\mathbf{\alpha}$ &
$\begin{bmatrix}0 & . & . \\ . & 1 & -5 \\ . & 0 & 0 \end{bmatrix}$& 
$\begin{bmatrix}-5 & -13 & . \\  9 &  -6 & . \\ . &   . & 0\end{bmatrix}$ &
$\begin{bmatrix} . & -8 & . \\   8 & . & . \\  . &   . & .\end{bmatrix}$
\\
& (unique axis $x$) & (std. setting) & (std. setting) \\
\end{tabular*}
\vskip 1mm
\begin{tabular*}{0.97\columnwidth}{@{\extracolsep{\fill}}rcccccc}
\hline
Fe $\rightarrow$ & Fe & (Fe,Cr) & {\sl Cr} & {\sl Mn} &
{\sl Co} & {\sl (Mn,Ni)}  \\ 
$\alpha_{\rm max}$($\hat{\alpha}_{\rm max}$) & 5(5)& 14(13)& 8(4)
& 15(11) & 6(2) & 3(3)
\\ [1mm]\hline\hline
\end{tabular*}
\end{table}

Table~\ref{tab1} summarizes our results and Fig.~\ref{fig1}(a) sketches the
${\cal E}$-induced spin canting that underlies the computed effects. The
obtained ME tensors satisfy symmetry relationships within the calculation
accuracy; e.g., for {\sl BiCrO$_3$} we got $\alpha_{12}=-\alpha_{21}$, as
expected for the $3m$ magnetic point group~\cite{borovik06}. Interestingly,
the largest ME coefficients obtained share the following two features: (i) The
applied ${\cal E}$ is perpendicular to ${\cal P}^{\rm S}$, which reflects the
{\sl easy-polarization-rotation} mechanism that determines the largest
electromechanical responses in ferroelectrics~\cite{fu00}. (ii) The change in
${\cal M}$ is perpendicular to the direction of the easy axis, which is a
signature of spin canting and an underlying spin-orbit mechanism. In contrast,
many symmetry-allowed coefficients were computed to be essentially null, which
reflects (i) the structural hardness of specific directions and (ii) the
energy cost of changing the magnetic moment along the direction of the easy
axis (which would typically require charge transfer between magnetic
ions~\cite{iniguez08}).

For all compounds, the full ME response is bigger than the frozen-cell effect,
the increase being significant in {\sl BiCrO$_3$}, {\sl BiMnO$_3$}, and {\sl
  BiCoO$_3$}. To illustrate how the strain-mediated response operates, we
consider the representative case of the $\alpha_{12}$ coefficient of {\sl
  BiCrO$_3$}. There, the largest strain response to ${\cal E}_1$ is associated
with the shear $\eta_5$, as quantified by the {\sl piezoelectric strain}
coefficient $d_{15}=e_{1j}(C^{-1})_{j5}=15$~pC/N. The resulting change in
${\cal M}_{2}$ is quantified by the piezomagnetic stress coefficient
$h_{25}$=$-$5$\times$10$^{4}$~A/m. This leads to a strain-mediated
contribution to the ME response of about $-$3$\times$10$^{-4}$~g.u.

Our calculations allow us to identify the atomic relaxation mechanisms that
mediate the ME response. Let us consider first the frozen-cell ME tensor
(Eq.~(\ref{eq:hatalpha})), whose lattice part can be expressed in terms of the
eigenmodes of $K$ as $\Omega_0^{-1} p^{\rm d}_{s\alpha}\kappa_{s}^{-1}p^{\rm
  m}_{s\mu}$. Here, $\kappa_s$ is the mode eigenvalue or {\sl stiffness}, and
$p^{\rm d}_{s\alpha}$ and $p^{\rm m}_{s\mu}$ are, respectively, the dielectric
and magnetic polarities introduced in Ref.~\onlinecite{iniguez08}. This
expression allows us to make a cumulative plot as the one shown in
Fig.~\ref{fig1}(b) for $\hat{\alpha}_{23}$ of BFO. Clearly, only two sets of
modes contribute significantly: (i) low-lying Bi-dominated modes with a
relatively small magnetic polarity ($\sim$78~\% of the response) and (ii)
high-energy Fe-dominated modes that involve a relatively large change in
${\cal M}$. We can thus conclude that the ME response is largely driven by Bi
modes that do not involve the magnetic species significantly, a result that
applies to all the studied compounds. The role of Bi is also preeminent in the
strain-mediated part of the response, the largest $B_{mj}$ couplings being
unequivocally associated with Bi.

The work of Ref.~\onlinecite{zvezdin06} allows us to compare theory and
experiment for BFO. These authors studied the bulk material under magnetic
fields up to 25~T, which allowed them to eliminate the spin cycloid and reach
a phase (G-type AFM) that resembles the one we have simulated. One can thus
extrapolate the ${\cal M}({\cal H})$ curve measured within the high-field
phase back to ${\cal H}$=0, and estimate the magnetization at zero field. The
result is 0.25~emu/g~=~0.028~$\mu_{\rm B}$/cell, in remarkable agreement with
our computed 0.036~$\mu_{\rm B}$/cell. Further, the ME response of the
high-field phase is about 4$\times$10$^{-4}$~g.u. at 10~K, again in excellent
agreement with our computed 5$\times$10$^{-4}$~g.u. While this comparison is
not fully justified, it certainly supports the physical soundness of our
results. Comparing our theory with measurements of BFO thin films
\cite{wang03} is unfortunately not possible, as the studied systems display
effects (e.g., very large net magnetic moments and possible presence of
Fe$^{2+}$) that clearly do not correspond to our simulations.

To put our results in perspective, note that the largest ME responses
measured for transition-metal compounds correspond to boracites, with
$\alpha_{\rm max}$=20$\times$10$^{-4}$~g.u. for Co$_3$B$_7$O$_{13}$Br
(see Table Table~1.5.8.2 of Ref.~\onlinecite{borovik06}). Larger
responses are observed in rare earth compounds, the greatest one being
$\alpha_{\rm
  max}$=100$\times$10$^{-4}$~g.u. for TbPO$_4$. It is important to
realize, though, that these maximum $\alpha$'s correspond to (very
low) temperatures slightly below the magnetic ordering transition
(17~K for Co$_3$B$_7$O$_{13}$Br and 2.2~K for
TbPO$_4$~\cite{borovik06}), where the ME effect is strongly enhanced.
In contrast, the response of BFO computed at
0~K is of the same order of magnitude as the $\alpha_{\rm max}$'s of
boracites, is expected to grow with $T$~\cite{fn:temperature}, and
should occur at $T_{\rm room}$ and above. Hence, when compared with
other magnetoelectrics, BFO is clearly an unique, very promising
material.

Interestingly, from a materials-design perspective one could say that BFO is
rather unsatisfactory, for several reasons. (1) The distortions mediating the
ME effect are dominated by non-magnetic ions, which is obviously not
ideal. (2) These materials are quite stiff, as reflected by the relatively
small piezoelectric and lattice-mediated dielectric responses: The largest
ones obtained were, respectively, $d_{\rm max}$$\approx$60~pC/N for {\sl
  BiMnO$_3$} and $\epsilon^{\rm latt}_{\rm max}$$\approx$45 for {\sl
  BiCoO$_3$}, while for prototype ferroelectric BaTiO$_3$ one gets $d_{\rm
  max}$$\approx$250~pC/N and $\epsilon^{\rm latt}_{\rm
  max}$$\approx$60~\cite{wu05}. (3) These materials are poor piezomagnets: For
the so-called {\sl piezomagnetic strain tensor} $g_{\mu j}$=$h_{\mu
  k}(C^{-1})_{jk}$ we obtained a maximum value of
0.4$\times$10$^{-10}$~Oe$^{-1}$ for {\sl BiCrO$_3$}, which is about one order
of magnitude smaller than what is typical in transition-metal compounds (see
Table~1.5.7.2 of Ref.~\onlinecite{borovik06}). Yet, in spite of all these
drawbacks, BFO is a competitive magnetoelectric.

If we were able to improve on some of these aspects, BFO might become
an excellent magnetoelectric. Our results indicate the ME response of
the studied compounds owes its relatively large value to the
structural response to electric fields. The secondary role of the
magnetic effects is clearly reflected in the fact that the magnitude
of $\mathbf{\alpha}_{\rm max}$'s (Table~\ref{tab1}) does not correlate
with the magnitude of the spin-orbit coupling of the considered
magnetic species, which rules out chemical substitution of iron as a
direct way to enhance the ME response. It seems more promising to try
to increase BFO's electromechanical responses, which might be
achieved, for example, in (i) BFO thin films strain-engineered to be
monoclinic (as opposed to rhombohedral), since monoclinicity is
usually accompanied by structural softness~\cite{fu00}, or (ii)
Bi$_{1-x}$La$_x$FeO$_3$ solid solutions where structural transitions
occur for small La concentrations~\cite{zalesskii03}. We hope our work
will stimulate detailed experimental studies of these and similar
systems, which our theory suggests constitute the best hope for
obtaining very large linear ME effects in BFO.

This is work funded by MaCoMuFi (STREP\_FP6-03321). It also received some
support from the Spanish Government (FIS2006-12117-C04-01, CSD2007-00041). Use
was made of the Barcelona (BSC-CNS) and Galicia (CESGA) Supercomputing
Centers.


\begin{thebibliography}{99}
\bibitem{fn:piezo} Piezomagnetism is forbidden for magnetic point
  groups $432$, $\bar{4}3m$, and $m\bar{3}m$, as well as for any group
  including time reversal or its combination with spatial
  inversion~\protect\cite{borovik06}. Thus, magnetically ordered
  ferroelectrics (broken time reversal; polar point group) are
  piezomagnets.
\bibitem{fn:addcomments} Our approach is equivalent to introducing
  {\sl reduced} magnetic variables in analogy with WVH. Note, though,
  that here we are concerned with spin magnetization, and that dealing
  with piezomagnetic effects associated to orbital magnetization might
  be more subtle.
\bibitem{fn:spacegroup} We found that having the easy axis along $x$ ($Bb'$
  magnetic space group) is marginally more stable than having it along $y$
  ($Bb$), by 0.01~meV/cell. ($x$ direction perpendicular to mirror plane.) We
  thus studied both cases, obtaining very similar ME responses. Here we only
  report on the first case.
\bibitem{fn:highspin} We restricted our investigation to the substitutions
  presenting an insulating high-spin state.
\bibitem{fn:temperature} In soft-mode ferroelectrics, the $T$ dependence of
  the electromechanical responses is driven by the {\sl softening} of the
  force ($K$) and elastic ($C$) constants. We expect BFO's lattice-mediated ME
  response to behave similarly.
\end{thebibliography}
\end{document}